# Effects of Solvent on Polymer Chain Dimensions:
# A Born-Green-Yvon Integral Equation Study

Mark P. Taylor and J.E.G. Lipson

Department of Chemistry

Dartmouth College, Hanover, New Hampshire 03755

**Abstract:** The equilibrium properties of a tangent-hard-sphere polymer chain in a hard sphere monomer solvent are studied using a Born-Green-Yvon integral equation in conjunction with a two-site solvation potential. The solvation potential is constructed using low density results for a hard-sphere trimer in a hard sphere solvent. The BGY equation has been solved for polymers of lengths up to 100 for a range of solvent densities and solvent diameters. The theory accurately describes the compression of the average polymer dimensions with increasing solvent density. The theory also accurately describes the variation in this compression as a function of the solvent diameter, predicting a maximum effect when the solvent diameter is two to three times larger than the polymer site diameter. Scaling exponents relating the polymer dimensions to chain length and solvent density are also obtained.

**I. INTRODUCTION**

The statistical properties of a polymer in solution are strongly dependent on the nature of the solvent [1,2]. In a so called "good" solvent the dimensions of the polymer are expanded relative to those of a non-interacting or ideal polymer while in a "poor" solvent the polymer dimensions are collapsed relative to the ideal or theta state. A flexible hard-sphere chain with site diameter $\sigma$ in a fluid of hard sphere monomers with diameter D provides a model for a polymer in a solvent. Even in this simple athermal system interesting solvent effects are observed [3-6]. For $D/\sigma \leq 1$ the hard sphere fluid represents a good solvent for the hard-sphere chain at all densities. However, for larger $D/\sigma$ Suen *et al.* [6] have recently reported evidence that a high density hard sphere fluid provides a poor solvent for a long hard-sphere chain. In any case, with increasing solvent density the average dimensions of the hard-sphere polymer are compressed. Furthermore, the magnitude of this compression effect is strongly dependent on the ratio $D/\sigma$.

One approach for studying a *n*-mer chain in solution is to introduce a *n*-site solvation potential which accounts for the interactions of the solvent with the polymer [5,7,8]. Approximating the *n*-site potential as a sum of two-site potentials allows one to map the polymer-solvent problem onto that of an isolated polymer whose segments interact via an density-dependent effective potential. In this work we introduce a two-site solvation potential and study the effects of solvent size and density on a hard-sphere-chain polymer using a Born-Green-Yvon (BGY) integral equation [9].

## II. THEORY

**A. Intramolecular Distribution Functions**

In this work we study the equilibrium structure of a single polymer molecule in a solvent. The polymer is modeled as a flexible chain of *n* identical hard-sphere monomers with diameter $\sigma$ which are connected by universal joints of bond length $\sigma$. The solvent consists of N hard spheres of diameter D and the solvent density is $\rho = N/V$ where V is the total system volume. The configurational properties of the polymer can be expressed in terms of a set of intramolecular site-site distribution functions $w_{ij}(r; \rho, D)$ which are related to the probability that two sites on the polymer, *i* and *j*, are separated by a distance $r = |\vec{r}_i - \vec{r}_j|$. The exact expressions for the $w_{ij}$ functions are quite formidable since they involve integrals over the complete set of the N solvent molecule coordinates. We note that this full problem is amenable to a BGY type integral equation treatment [9,10] although here we take a much simpler tack.

A formal simplification in the definition of the $w_{ij}$ functions can be achieved by writing the integrals over the solvent coordinates in terms of an *n*-site solvation potential [7]. Now, by making the assumption that the *n*-site solvation potential can be written as a sum of two-site solvation potentials (Chandler and Pratts' superposition approximation) we can map the full problem described above on to that of an isolated chain with an modified site-site potential which implicitly accounts for the effects of the solvent. The resulting density dependent effective potential between nonbonded polymer sites $i$ and $j$ ($|i-j|>1$) is given by

$$u_{ij}^{\text{eff}}(r; \rho, D) = u_{ij}^{\text{hs}}(r) + u_{ij}^{\text{sol}}(r; \rho, D) \qquad (1)$$

where $u_{ij}^{\text{hs}}(r)$ is the original hard sphere potential

$$u_{ij}^{\text{hs}}(r) = \begin{cases} \infty & r < \sigma \\ 0 & r > \sigma \end{cases} \qquad (2)$$

and $u_{ij}^{\text{sol}}(r; \rho, D)$ is the site-site solvation potential.

The site-site distribution function for an $n$-mer chain whose segments interact via such an effective potential is given by

$$w_{ij}(r; \rho, D) = \frac{1}{Z_n(\rho)} \int \cdots \int \prod_{a=1}^{n-2} \prod_{b=a+2}^{n} \exp[-\beta u_{ab}^{\text{eff}}(r; \rho, D)] \prod_{\alpha=1}^{n-1} s_{\alpha,\alpha+1} \prod_{m \neq i,j}^{n} d\vec{r}_m \qquad (3)$$

where $\beta = 1/k_B T$, $s(r) = \delta(r-\sigma)/4\pi\sigma^2$ is the intramolecular distribution function between bonded sites, and $Z_n(\rho)$ is the single chain partition function, which ensures the normalization condition

$$\int dr\, 4\pi r^2\, w_{ij}(r; \rho, D) = 1. \qquad (4)$$

Due to the constraints of chain connectivity the $w_{ij}(r; \rho, D)$ functions are identically zero for $r > |i-j|\sigma$. A number of equilibrium configurational properties of the polymer chain may be expressed using averages over these distribution functions. In particular, the mean-square distance between sites $i$ and $j$ is

$$\langle r_{ij}^2 \rangle = \int dr\, r^2\, 4\pi r^2\, w_{ij}(r; \rho, D) \qquad (5)$$

and the mean-square radius of gyration is

$$\langle R_g^2 \rangle = \frac{1}{n^2} \sum_{i<j}^{n} \langle r_{ij}^2 \rangle. \qquad (6)$$

**B. BGY Integral Equation**

As we have shown previously [9], an integral equation for the intramolecular site-site distribution functions of an isolated interaction-site polymer can be obtained following the method of Born, Green, and Yvon [11,12]. The BGY equation for the end-to-end distribution function of an *n*-mer chain is given by

$$\vec{\nabla}_1 w_{1n} = \vec{\nabla}_1 [-\beta u_{1n}^{\text{eff}}(r; \rho, D)] w_{1n} + \int d\vec{r}_2 \, \vec{\nabla}_1 [\ln s_{12}] \, w_{12n}$$

$$+ \sum_{\nu=3}^{n-1} \int d\vec{r}_\nu \, \vec{\nabla}_1 [-\beta u_{1\nu}^{\text{eff}}(r; \rho, D)] \, w_{1\nu n} \qquad (7)$$

where $w_{1\nu n}$ is a three-site distribution function which satisfies the reduction condition

$$w_{1n} = \int d\vec{r}_\nu \, w_{1\nu n} . \qquad (8)$$

In order to solve the above BGY equation we must introduce two approximations. First, the required three-site functions are expressed in terms of two-site functions using a superposition-like approximation and, second, the distribution function between any two sites *i* and *j* is approximated by the end-to-end distribution function of a chain of length *i+j*–1. With these approximations, Eq. (7) can be solved recursively by direct numerical integration for a chain of arbitrary length. For details of the numerical solution the reader is referred to Ref. 9.

This BGY theory has been used to study both isolated hard-sphere and square-well chain and ring polymers [9,13]. The success of the theory in these cases provides validation of the above approximations. In this work we apply the BGY theory to study a polymer in solution via the solvation potential approach described in the previous section. The construction of the required solvation potential is described forthwith.

**C. The Solvation Potential**

Here we construct a polymer solvation potential which is motivated by low density results for a hard-sphere 3-mer in a hard sphere solvent. This task is accomplished by equating a low density expression for the hard-sphere 3-mer end-to-end distribution function to the analogous distribution function for an isolated 3-mer whose sites interact via an effective potential.

To first order in solvent density, the end-to-end distribution function for a hard-sphere 3-mer is given by

$$w_{13}(r; \rho, D) = \frac{\exp[-\beta u_{13}^{hs}(r)]}{Z_3(\rho)} [w_{13}^{(0)}(r) + \rho\, w_{13}^{(1)}(r; D)] \qquad \sigma \leq r \leq 2\sigma \qquad (9)$$

where the zeroth order coefficient is

$$w_{13}^{(0)}(r) = \int d\vec{r}_2\, s_{12}\, s_{23} = \frac{1}{8\pi\sigma^2 r} \qquad \sigma \leq r \leq 2\sigma \qquad (10)$$

and we write the first order coefficient as

$$w_{13}^{(1)}(r; D) = A^3\, w_{13}^{(0)}(r) [\delta_3(r; D) + \delta_4(\sigma, \sigma, r; D)]. \qquad (11)$$

The $f$-bond diagrams comprising this first order coefficient are defined as

$$\delta_3(r_{13}; D) = \int d\vec{r}_4\, f_{14} f_{34} = \frac{\pi}{12}(16X^3 - 12rX^2 + r^3) \qquad \sigma \leq r \leq 2X \qquad (12)$$

and

$$\delta_4(r_{12}, r_{23}, r_{13}; D) = \int d\vec{r}_4\, f_{14} f_{24} f_{34} \qquad (13)$$

where $X = (\sigma + D)/2$ and the solvent-polymer site interaction Mayer $f$-function is given by

$$f(r) = \begin{cases} -1 & r < X \\ 0 & r > X \end{cases} \qquad (14)$$

An analytic expression for $\delta_4$ is provided in Ref. 14. While rigorously the scaling factor A appearing in Eq. (11) should be unity, here we define it as the ratio of the solvent diameter

D to the interaction length X (i.e., A=D/X). In Table 1 we compare results for the mean-square end-to-end distance of a hard-sphere 3-mer in a $D=\sigma$ hard sphere solvent computed using the Eq. (9) first order expansion with those from Monte Carlo (MC) simulation [5]. The expansion correctly yields the solvent induced compression with increasing solvent volume fraction $\eta = \pi\rho D^3/6$, although the magnitude of this effect is underestimated.

The end-to-end distribution function of an isolated 3-mer whose sites interact via an effective potential is given exactly [15] by

$$w_{13}(r; \rho,D) = \frac{\exp[-\beta u_{13}^{eff}(r; \rho,D)]}{Z_3(\rho)} w_{13}^{(0)}(r) \ . \qquad \sigma \leq r \leq 2\sigma \qquad (15)$$

Using the definition of $u^{eff}(r)$ given in Eq. (1) and equating the above two expressions for $w_{13}(r)$ [Eqs. (9) and (15)] yields the following solvation potential

$$\beta u^{sol}(r; \rho,D) = -\ln\left[ 1 + \rho\, w_{13}^{(1)}(r; D)/w_{13}^{(0)}(r)\right] \ . \qquad \sigma \leq r \leq 2\sigma \qquad (16)$$

In Fig. 1 we plot this site-site solvation potential both as a function of solvent volume fraction $\eta$ and solvent diameter D. In all cases the potential is short ranged and purely attractive. For a given solvent diameter the strength of the potential increases with increasing solvent density. The range of the potential decreases with decreasing D and the potential vanishes in both the limits of D→0 and D→∞. In the following we apply this solvation potential to the problem of a hard-sphere polymer in a hard sphere solvent. This potential mimics a very local, solvent induced site-site interaction mediated by a single solvent molecule. The potential to some extent incorporates intramolecular screening effects and contains information about the solvent geometry. The potential does not include any information about the local solvent structure.

## III. RESULTS

Using the single chain BGY equation [Eq. (7)] and the above polymer solvation potential [Eq. (16)] we have computed the intramolecular distribution functions of hard sphere chains in a hard sphere solvent for chains of lengths $4 \leq n \leq 100$, solvent diameters $0.1 \leq D/\sigma \leq 100$, and solvent volume fractions $0.0 \leq \eta \leq 0.5$ ($\eta = \pi \rho D^3/6$). Rather than show the actual $w_{ij}(r; \rho, D)$ functions, here we present averages over these distribution functions in the form of the polymer mean-square end-to-end distance and radius of gyration defined in Eqs. (5) and (6), respectively.

In Fig. 2 we show the variation in average polymer size with solvent density. BGY results for the mean-square end-to-end distance are shown for chains of lengths $n$=10, 20, and 30 and solvent diameter $D=\sigma$. Also shown in this figure are MC data for the polymer solvent system [4,5] and, in the case of $n$=20, MC data for an analogous polymer melt system [16]. The BGY theory is seen to correctly predict that the polymers are moderately compressed with increasing solvent density. For isolated chains (i.e., $\eta$=0.0) the BGY theory slightly overestimates $\langle r_{1n}^2 \rangle$ and this discrepancy is seen to persist with increasing solvent density. Overall the BGY theory gives quite good agreement with the MC data even up to very high volume fractions. Note from Fig. 2 that both the average size, and its variation with density, of a hard-sphere 20-mer in a melt of hard-sphere 20-mers (filled symbols) is quite similar to the corresponding results for a hard-sphere 20-mer in a solvent of hard-sphere 1-mers.

For a long chain in a solvent one expects a scaling relationship of the form $R^2 \sim n^{2\nu}$ (where $R^2$ is either the mean-square end-to-end distance or radius of gyration), where an exponent of $2\nu = 1$ corresponds to a theta solvent while $2\nu > 1$ and $2\nu < 1$ correspond to good and poor solvents, respectively [1,2]. We have computed this exponent for hard-sphere chains in a hard sphere solvent with $D=\sigma$. Our BGY results for $\langle R_g^2 \rangle$, for $20 \leq n \leq 100$, give exponents of $2\nu = 1.27$, 1.21, and 1.16 for solvent volume fractions of $\eta = 0.0$, 0.3, and 0.5, respectively. The fact that the $\eta$=0.0 result exceeds the Flory value of 6/5 indicates that the chain lengths considered here are not large enough to be in the asymptotic scaling regime. However, we can still conclude that the D=σ solvent is a "good" solvent for all densities and that increasing solvent density corresponds to making the solvent conditions more "theta-like". This latter finding suggests that an athermal 1-mer solvent produces similar effects on the dimensions of a single $n$-mer chain as an athermal $n$-mer solvent (for which $2\nu \rightarrow 1$ with increasing system density [17]).

For long chains in a dense solvent one also expects the scaling relationship $R^2 \sim \eta^{-\gamma}$ with $\gamma \approx 0.25$ for a good solvent [18]. Although this value of $\gamma$ presumes that one is in the ideal scaling regime (i.e., $2\nu=1$), we have nevertheless computed it for our hard-sphere chain/solvent system. Our BGY results for a hard-sphere $n$-mer in a hard sphere solvent, with D=σ and $0.3 \leq \eta \leq 0.5$, give exponents of $\gamma \approx 0.3$ and 0.4 for $n$=50 and 100, respectively. The MC melt data, shown in Fig. 2, give a scaling exponent of $\gamma$=0.24 ($\eta \geq 0.2$) while the BGY results for the polymer solvent system with $n$=20 give $\gamma$=0.26.

In Fig. 3 we show both BGY and MC results for the mean-square end-to-end distance as a function of solvent diameter D ($0.1 \leq D/\sigma \leq 100$), for a hard-sphere 30-mer in a hard sphere solvent at solvent volume fraction $\eta = 0.2$. The BGY theory predicts that in the limits of a very large and very small solvent diameter D/σ the average polymer size is essentially unperturbed from the isolated chain result. (We note that the average size of an isolated hard-sphere 30-mer is approximately equivalent to a sphere of diameter 7σ). When the solvent diameter D is similar to the polymer site diameter σ the polymer is compressed by the solvent with the maximum effect occurring for D/σ ~ 2 to 3. In comparison with the Monte Carlo results [4], the BGY theory is found to be quite accurate.

## IV. CONCLUSION

In this work we have studied the equilibrium properties of a hard-sphere polymer chain in a hard sphere monomer solvent. The polymer-solvent system is treated in the context of a polymer site-site solvation potential.  Using a BGY integral equation we have computed the intramolecular distribution functions for polymers over a range of solvent diameters and densities.  The BGY theory accurately describes the compression of the average polymer dimensions with increasing solvent density.  Additionally, the BGY theory accurately describes the variation in this compression as a function of solvent diameter, predicting a maximum effect when the solvent diameter is two to three times larger than the polymer site diameter

**Acknowledgments**

Financial support from the National Science Foundation (grant DMR-9424086) and the Camille and Henry Dreyfus Foundation is gratefully acknowledged.

## References

[1]   P.J. Flory, Principles of Polymer Chemistry, Cornell University, Ithaca, 1953.

[2]   A.Yu. Grosberg and A.R. Khokhlov, Statistical Physics of Macromolecules, AIP, New York, 1994.

[3]   J. Komorowski and W. Bruns, J. Chem. Phys., 103 (1995) 5756-5761.

[4]   F.A. Escobedo and J.J. de Pablo, Mol. Phys., 89 (1996) 1733-1754.

[5]   C.J. Grayce, J. Chem. Phys., 106 (1997) 5171-5180.

[6]   J.K.C. Suen, F.A. Escobedo, and J.J. de Pablo, J. Chem. Phys., 106 (1997) 1288-1290.

[7]   D. Chandler and L.R. Pratt, J. Chem. Phys., 65 (1976) 2925-2940.

[8]   C.J. Grayce and K.S. Schweizer, J. Chem. Phys., 100 (1994) 6846-6856.

[9]   M.P. Taylor and J.E.G. Lipson, J. Chem. Phys., 104 (1996) 4835-4841.

[10]  M.P. Taylor and J.E.G. Lipson, J. Chem. Phys., 102 (1995) 2118-2125.

[11]  M. Born and H.S. Green, Proc. Roy. Soc., London, Ser. A, 188 (1946) 10-18.

[12]  J. Yvon, Actual. Sci. Ind., 203 (1935).

[13]  M.P. Taylor, J.L. Mar, and J.E.G. Lipson, J. Chem. Phys., 106 (1997) 5181-5188.

[14]  M.J.D. Powell, Mol. Phys., 7 (1964) 591-592.

[15]  M.P. Taylor, Mol. Phys., 86 (1995) 73-85.

[16]  A. Yethiraj and C.K. Hall, J. Chem. Phys., 96 (1992) 797-807.

[17]  P.J. Flory, J. Chem. Phys., 17 (1949) 303-310.

[18]  M.K. Kosmas and K.F. Freed, J. Chem. Phys., 69 (1978) 3647-3659.

Table 1.  Mean-square end-to-end distance for a tangent hard sphere 3-mer in a hard sphere solvent (D=σ) computed via a first order density expansion and via Monte Carlo simulation [5].

| η    | ρ-expansion | Monte Carlo |
|------|-------------|-------------|
| 0.0  | 2.50        | 2.50        |
| 0.1  | 2.48        | 2.47        |
| 0.2  | 2.45        | 2.44        |
| 0.3  | 2.43        | 2.40        |
| 0.4  | 2.41        | 2.36        |
| 0.45 | 2.40        | 2.34        |

**Figure Legends**

Fig. 1. Solvation potential as a function of site-site separation r over a range (a) of solvent volume fractions η with D=σ and (b) of solvent diameters D with η=0.4.

Fig. 2. Variation of the mean-square end-to-end distance with solvent density η for a hard-sphere *n*-mer chain in a hard sphere solvent of diameter D=σ. Solid lines are the results of the BGY theory and the open symbols are MC data from Gracye (◊) [5] and from Escobedo and de Pablo (O, □) [4]. The filled symbols are MC data for an *n*=20 polymer melt system taken from Yethiraj and Hall [16].

Fig. 3 Variation of the mean-square end-to-end distance with solvent diameter D for a hard-sphere 30-mer chain in a hard sphere solvent of volume fraction η=0.2. Solid lines are the results of the BGY theory and the symbols are MC data from Escobedo and de Pablo [4]. The dashed line indicates the BGY result for an isolated hard-sphere 30-mer.

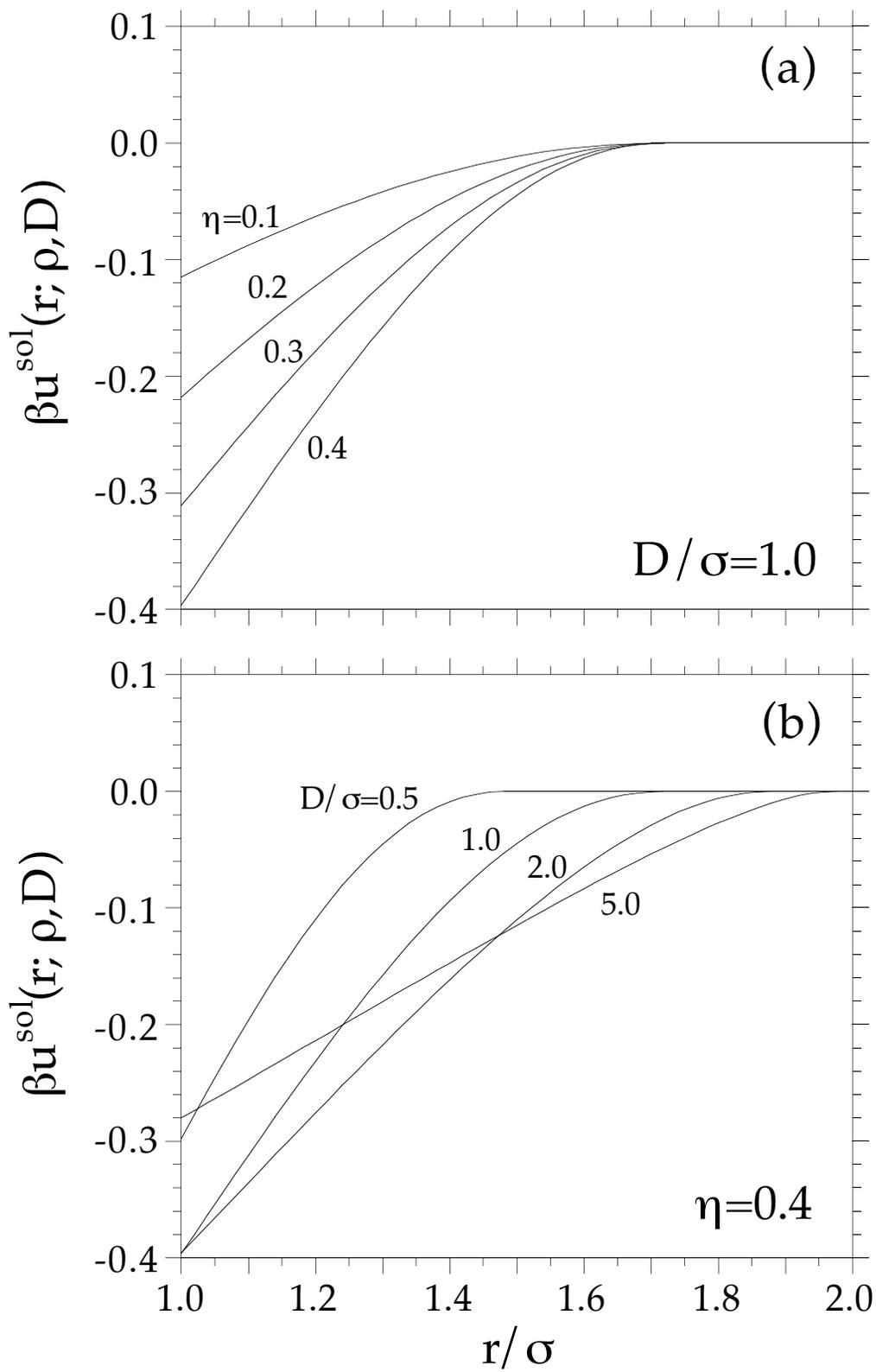

Taylor and Lipson - Fig. 1

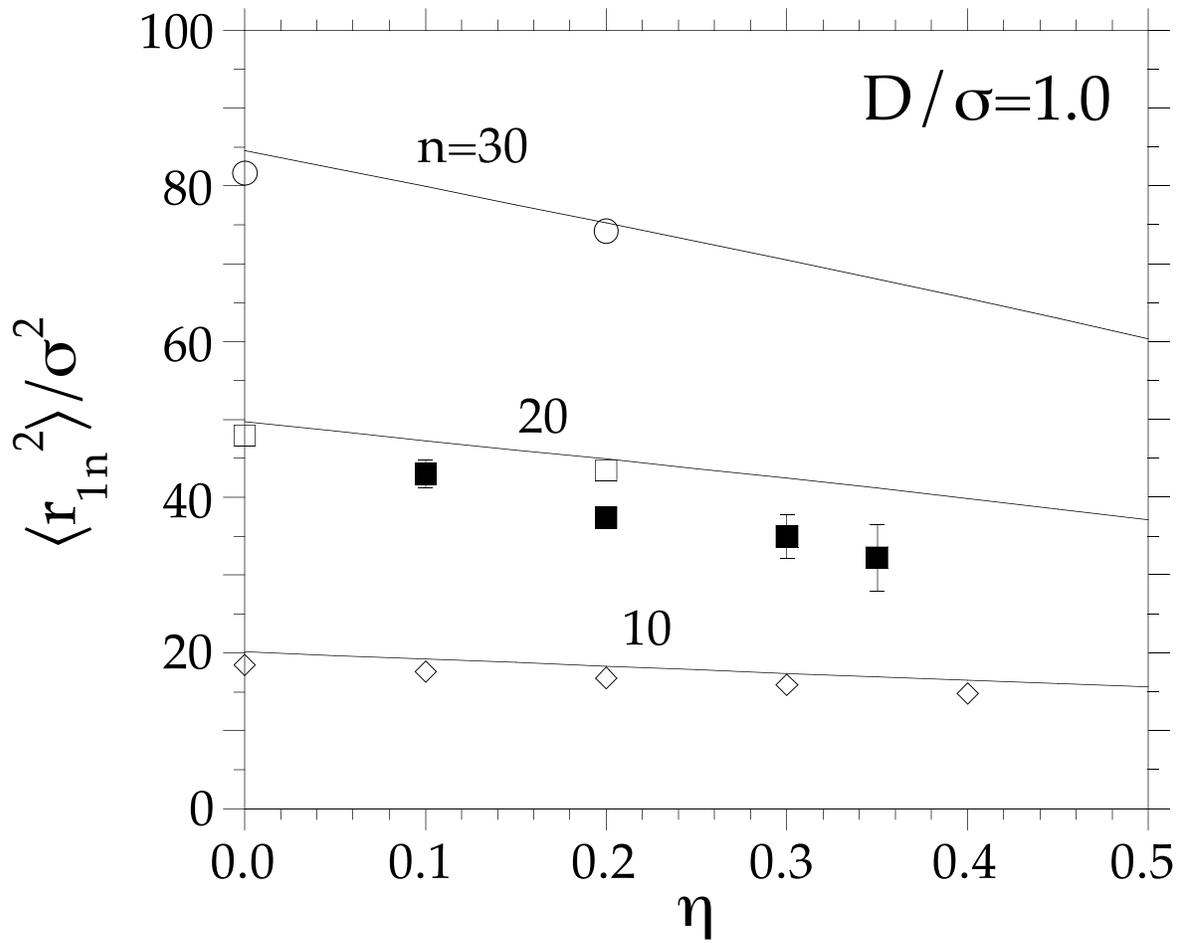

Taylor and Lipson - Fig. 2

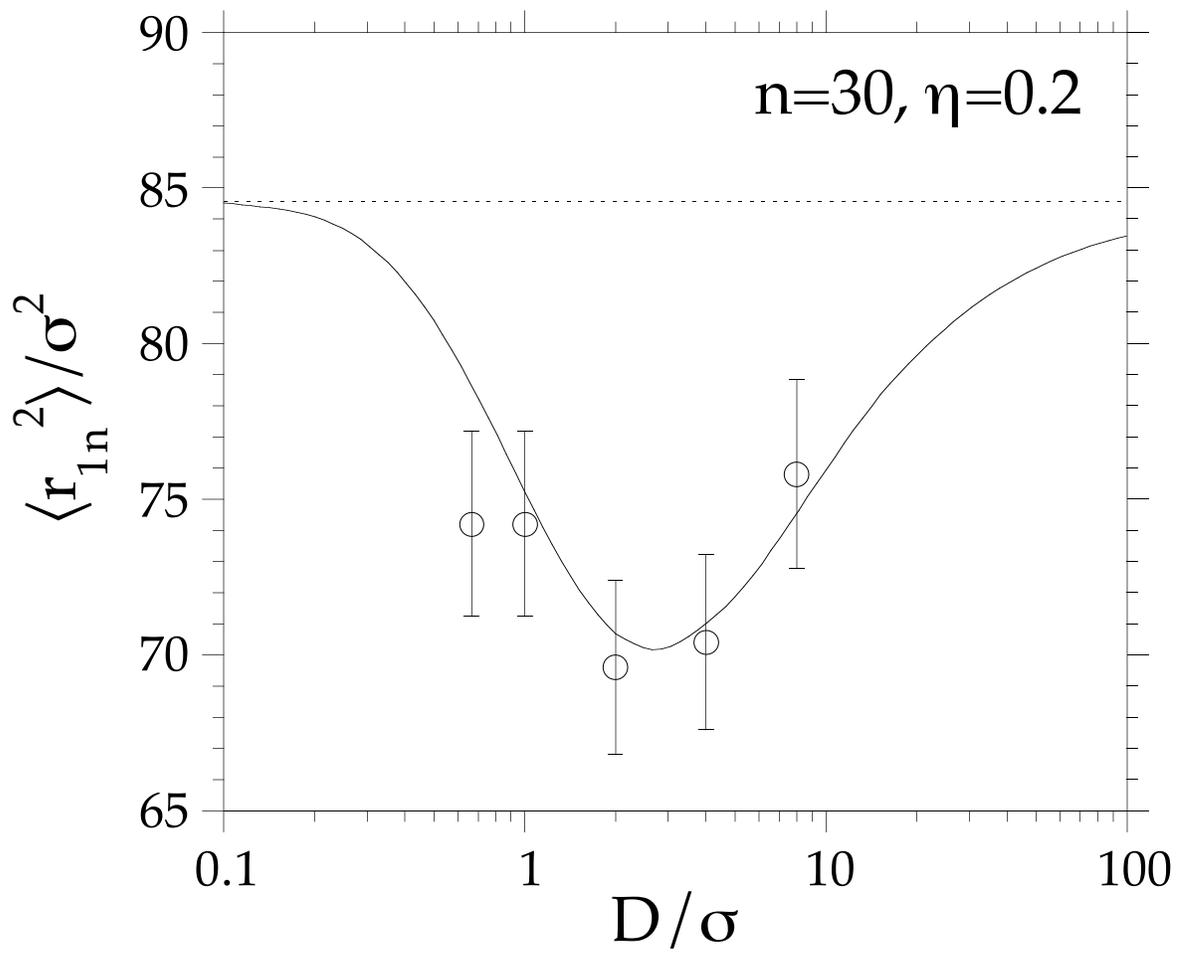

Taylor and Lipson - Fig. 3